\documentstyle[preprint,aps]{revtex}

\newcommand{\ba}{\begin{array}}
\newcommand{\ea}{\end{array}}

\newcommand{\ger}{\stackrel{\scriptstyle >}{\scriptstyle\sim}}

\def\bC{\smash{\overline C}\vphantom{C}}

\def\bphi{\smash{\overline \phi}\vphantom{\phi}}
\newcommand{\bn}{\overline{n}}
\newcommand{\bm}{\overline{m}}

\newcommand{\bq}{\overline{q}}

\newcommand{\bl}{\overline{l}}

\newcommand{\brho}{\overline{\rho}}
\newcommand{\bdelta}{\overline{\delta}}
\newcommand{\bsigma}{\overline{\sigma}}

\newcommand{\tmu}{\tilde{\mu}}
\newcommand{\tK}{\tilde K}
\newcommand{\tY}{\tilde Y}
\newcommand{\tf}{\tilde f}

\newcommand{\hK}{\hat K}
\newcommand{\hW}{\hat W}
\newcommand{\albe}{\alpha\beta}
\newcommand{\gbd}{\gamma {\overline{\delta}}}
\newcommand{\abgbd}{\alpha\beta\gamma {\overline{\delta}}}
\newcommand{\abg}{\alpha\beta\gamma}
\newcommand{\loopf}{\frac{\Lambda^2}{16\pi^2}}
\newcommand{\loopff}{\left(1+\frac{(N_c-5)\Lambda^2}{16\pi^2}\right)}
\def\bea{\begin{eqnarray}}
\def\eea{\end{eqnarray}}

\def\al{\alpha}

\def\be{\begin{equation}}
\def\ee{\end{equation}}
\def\bea{\begin{eqnarray}}
\def\eea{\end{eqnarray}}
\def\ap{\approx}

\begin{document}

\preprint{
FTUAM 97/10
\hspace{-37mm}
\raisebox{2.5ex}{
KAIST-TH 97/13}}

\title{\bf Supergravity Radiative Effects on Soft Terms and the $\mu$ Term} 
\author{Kiwoon CHOI,
Jae Sik LEE 
and Carlos MU\~NOZ \thanks{Permanent address:
Departamento de
F\'{\i}sica Te\'orica, Universidad Aut\'onoma de Madrid, 
Cantoblanco, 28049 Madrid, Spain}}
\address{
Department of Physics,
Korea Advanced Institute of
Science and Technology,
\\
Taejon 305-701, Korea}

\maketitle

\begin{abstract}
We compute quadratically divergent
supergravity one--loop effects 
on  soft supersymmetry--breaking parameters
and the $\mu$ term in generic hidden sector supergravity models.
These effects can significantly modify 
the matching condition for soft parameters at the Planck scale
and also provide several new sources of the $\mu$ term which
are naturally of order the weak scale.
We also discuss some phenomenological implications of these effects,
particularly  the violation of the scalar mass universality
which may lead to dangerous FCNC phenomena, and apply the results 
to superstring effective supergravity models.

\end{abstract}


\narrowtext

Supergravity (SUGRA) models with a   supersymmetry (SUSY)--breaking
hidden sector 
provide an attractive theoretical scheme 
to explain the soft SUSY--breaking terms in the SUSY standard model \cite{BIM3}.
If SUSY is broken at a scale  
$M_S$  
in  such models,
soft parameters have a scale of order 
$m_{3/2}\ap M_S^2/M_P$ 
where 
$m_{3/2}$ is the gravitino mass and
$M_P\ap 2.4\times 10^{18}$ GeV
denotes the reduced Planck mass.
One then obtains the desired weak-scale soft terms if $M_S$
has an intermediate scale value, $M_S\ap 10^{10}$ GeV.
Besides the soft parameters, the SUSY standard model contains 
another mass scale, the Higgsino mass $\mu$.
Understanding why $\mu$ has the weak scale value, not
$M_P$ for instance, has been
often called the $\mu$ problem.
It has been known that hidden sector SUGRA models can
accommodate mechanisms which lead to $\mu\ap m_{3/2}$
in a natural manner, thereby solving the $\mu$ problem \cite{BIM3}.

Usual computation of the soft parameters and $\mu$  in 
SUGRA models goes as follows. One first computes these parameters
at the {\it tree level} of
SUGRA interactions \cite{SW}.
The results are then interpreted as a  matching condition  
for the soft terms and the $\mu$ term  at $M_P$ 
which corresponds to the scale
of the messenger SUGRA interactions.
Once the values at $M_P$ are given by this tree--level
matching condition,
one takes account of the logarithmic running
due to the renormalizable
gauge and Yukawa interactions  in order to find the value of the 
parameters  
at low--energy scales. 

However,
there can be a significant modification
of the matching condition in this procedure 
due to the quantum corrections induced
by non-renormalizable SUGRA interactions \cite{KR}.
These SUGRA radiative  corrections  are quadratically
divergent and thus they do not affect the logarithmic running,  
but alter the matching condition at the cutoff scale
$\Lambda\ap M_P$.
Note that  essentially this amounts  
to the modification of the  messenger SUGRA interactions and thus $\Lambda$ has
to be taken as the messenger scale $M_P$ (or the string
or compactification scale in string theory), 
{\it not} for instance
the intermediate SUSY breaking scale $M_S$.
In this paper, we wish to examine the SUGRA radiative effects
on the soft parameters and $\mu$ at $M_P$  
in generic hidden sector  SUGRA models,
and discuss their implications.
As we will see, 
the modifications due
to quadratically divergent SUGRA effects are qualitatively
different from those due to the logarithmic running effects.
For instance, even when the tree--level matching condition gives a universal
soft scalar mass at $M_P$,  
SUGRA radiative effects  can significantly {\it spoil}  
the  universality, 
which may lead to dangerous flavor changing neutral current (FCNC)
phenomena  at low energies.
They also provide several {\it new} sources of the $\mu$-term
which are naturally of order the weak scale.
Taking account of SUGRA radiative effects
will be  particularly important in the case
that the tree--level matching condition
gives vanishing (or significantly smaller than $m_{3/2}$)
soft parameters and/or $\mu$ at $M_P$.
In this case, one--loop SUGRA effects can provide 
a leading contribution,
{\it not} merely a 
subleading correction.
The above aspects  will be discussed later in the context of
superstring effective supergravity models. 


Our starting point is the  bosonic part of the SUGRA Lagrangian related with the
scalar fields, including the
relevant one-loop  correction \cite{gaillard}: 
\begin{eqnarray}
g^{-1/2}{\cal L} &&= 
\left[K_{I\overline J}      
+\frac{{\Lambda}^2}{16\pi^2}
(R_{I\overline{J}}-
2K_{I\overline{J}})
\right]\partial_{\mu}
\phi^I\partial^{\mu}\bphi^{\overline J}
\nonumber\\ 
&&
-\left[1+
\frac{{\Lambda}^2}{16\pi^2}
(N_c-5)\right]
e^G \left(G_I K^{I{\overline J}} G_{\overline J}-3
\right)
\nonumber\\
&&-\
\frac{{\Lambda}^2}{16\pi^2}
e^G\left[\left(N_c-1\right)-K^{I{\overline{L}}}G_{\overline L}
R_{I\overline{J}}
G_M K^{M{\overline J}} \right]
\ ,
\label{F2}
\end{eqnarray}
where the K\"ahler function 
$G=
K+\log|W|^2$
is  a combination of the real K\"ahler
potential $K(\phi^I,\bphi^{\overline I})$ 
and the holomorphic superpotential
$W(\phi^I)$,
$G_I\equiv \partial_I G\equiv
\partial G/\partial\phi^I$, the matrix $K^{I{\overline J}}$ is
the inverse of the K\"ahler metric 
$K_{I{\overline J}}\equiv  
\partial_I
\partial_{\overline{J}}K$, and finally
$N_c$ denotes the total number of chiral multiplets in the model.
To simplify the notation,
we have introduced a generalized Ricci
tensor
\be
R_{I\overline{J}} = \partial_I
\partial_{\overline{J}}\ln [{\rm det}(K_{L\overline{M}})/
{\rm det}({\rm Re}f_{ab})]\ ,
\label{F17}
\ee 
where $f_{ab}(\phi^I)$ is the holomorphic gauge kinetic function,
with $a, b$ denoting gauge indices.
Here we ignore the $D$-part of the scalar potential 
with the assumption that SUSY is broken by the $F$-components
of some hidden sector fields. The pieces 
proportional to $\Lambda^2$
represent the quadratically divergent SUGRA one--loop corrections
which will modify the matching conditions for the soft parameters and 
$\mu$ at $\Lambda\ap M_P$. Note that the logarithmically divergent
pieces can be properly taken into account by the well known  renormalization
group running.
Unless explicitly specified, we will use throughout this paper 
the standard SUGRA 
mass unit in which $M_P=1$, and thus for instance $\Lambda$ denotes
the cutoff scale in this unit.

In hidden sector SUGRA models, it is convenient to split the chiral
multiplets into two categories, the hidden sector multiplets
$\{h^m\}$ 
and the observable
sector multiplets $\{C^{\al}\}$ 
\cite{foot1}. 
The hidden sector fields develop large vacuum expectation
values (VEVs) $\ap M_P$ and are responsible
for SUSY breaking if some of their auxiliary components  
develop non--vanishing VEVs.
On the other hand, 
the observable sector fields
have the VEVs $\ll M_P$, allowing 
an expansion of 
$K$, $W$ and $f_{ab}$
in powers of $C^{\al}$:
\begin{eqnarray}
K &&=
{\hat K}
+ {\tilde K}_{{\alpha}{\overline{\beta}}}
C^{\alpha} { \bC^{\overline{\beta}} }\ 
+\frac{1}{4}
{\tilde K}_{{\alpha}{\overline{\beta}}\gamma{\overline {\delta}}}
C^{\alpha} \bC^{\overline{\beta}}
C^{\gamma} \bC^{\overline{\delta}}
\nonumber\\ 
&&+
\left[ \frac{1}{2} Z_{{\alpha }{ \beta }}
C^{\alpha}C^{\beta } 
+\frac{1}{2}
{Z}_{{\alpha}\beta{\overline{\gamma}}}
C^{\alpha} C^{\beta} \bC^{\overline{\gamma}} \right.
\nonumber \\
&& 
+ \left. \frac{1}{6} Z_{{\alpha }{ \beta }\gamma\overline{\delta}}
{C}^{\alpha}
C^{\beta } C^{\gamma} \bC^{\overline{\delta}} 
+\ h.c.   \right]
+...\ , 
\nonumber 
\\
W &&= {\hat W} +\frac{1}{2}\tilde{\mu}_{{\alpha}{\beta}}{C}^{\alpha}C^{\beta} 
+ \frac{1}{6} \tilde{Y}_{{\alpha}{\beta}{\gamma}}
{C}^{\alpha}C^{\beta}C^{\gamma} 
+... \ ,
\nonumber 
\\
f_{ab} &&= 
\hat{f}_{ab}+\frac{1}{2}{\tilde f}_{ab\alpha\beta}C^{\alpha}C^{\beta}
+... \ ,
\label{F4}
\end{eqnarray}
where the ellipsis denote the terms 
which are irrelevant for the present calculation.
Note that the coefficient functions in $K$ depend upon 
both $h^m$ and $\overline{h}^{\overline m}$, while the coefficient functions
in $W$ and $f_{ab}$ are holomorphic functions of $h^m$ only.
It is well known that the quadratic terms associated with
$Z_{{\alpha}{\beta}}$ and 
$\tilde{\mu}_{{\alpha}{\beta}}$ 
are relevant for the $\mu$ term  \cite{BIM3}.
As we will discuss below, when SUGRA corrections are included, 
the cubic and quartic terms associated with   
${Z}_{{\alpha}\beta{\overline{\gamma}}}$ and
$Z_{{\alpha }{ \beta }\gamma\overline{\delta}}$, 
and  the quadratic terms associated with
${\tilde f}_{ab\alpha\beta}$ are also relevant for $\mu$.

Then integrating out the hidden sector fields in
the Lagrangian (\ref{F2}),
we find the kinetic terms and 
the 
soft SUSY--breaking scalar potential of the observable
sector fields  as
\bea
{\cal L}_{\rm eff}&&=
K^{\prime}_{\alpha\overline{\beta}}
\partial_{\mu}C^{\alpha}
\partial^{\mu}{\bC}^{{\overline {\beta}}}
-{m}'^2_{{\alpha}{\overline {\beta}}}
C^{\alpha} {\bC}^{\overline{\beta}}
\nonumber
\\
&&-[  
\frac{1}{2} B'_{\alpha \beta} C^{\alpha}C^{\beta} 
+\frac{1}{6} A'_{\alpha \beta \gamma} 
	    {C}^{\alpha}{C}^{\beta}{C}^{\gamma}
+ h.c. ]\ , 
\label{F7}
\eea
where the kinetic coefficients are  given by
\be
K^{\prime}_{\alpha\overline{\beta}} =
\tilde{K}_{\alpha {\overline{\beta}}}
+\frac{{\Lambda}^2}{16\pi^2}
(R_{\alpha\overline{\beta}}-
2{\tilde K}_{\alpha\overline{\beta}})\ ,
\label{kco}
\ee
and the primes in soft parameters mean that they 
are defined for {\it un-normalized} fields.
The  soft masses and $A$-coefficients are  
\begin{eqnarray}
{m}'^2_{{\alpha}{\overline{\beta}}} &&=
V_1 {\tilde K_{{\alpha}{\overline{\beta}}}}+
\left(m_{3/2}^2{\tilde K_{{\alpha}{\overline{\beta}}}}
- {F}^{m} R_{m{\overline n}\alpha {\overline {\beta}}}
{\overline F}^{\overline n}\right)\loopff
\nonumber\\ 
&&-\loopf
\left[
m_{3/2}^2 R_{{\alpha}{\overline{\beta}}} 
+ m_{3/2} F^m D_m
R_{{\alpha}{\overline{\beta}}}
+ m_{3/2}{\overline F}^{\overline n} D_{\overline n}
R_{{\alpha}{\overline{\beta}}} \right.
\nonumber \\
&&\left. +F^m \left( D_m D_{\overline n}
R_{{\alpha}{\overline{\beta}}}
- R_{\overline n}^{\overline l}
 R_{m {\overline l}{\alpha}{\overline{\beta}}}
-R^l_mR_{l\overline{n}\alpha\overline{\beta}}
+R^{\gamma}_{\alpha}R_{m\overline{n}\gamma\overline{\beta}}
\right){\overline F}^{\overline n} \hspace{0.05 cm}\right]
\ ,
\label{F8}
\\
A^{\prime}_{\abg} &&= 
e^{\hK/2}\frac{\hW^*}{|\hW|}\left\{
F^m \left(\tY_{\abg} \partial_m\hK+D_m \tY_{\abg}\right) \loopff \right.
\nonumber \\
&& -\loopf \left[
F^m R_m^n \left(\tY_{\abg} \partial_n\hK+D_n \tY_{\abg}\right) 
+ \tK^{\delta\brho}F^m\left(D_mR_{\brho ( \alpha }\right)
\tY_{\beta\gamma ) \delta} \right.
\nonumber \\
&&-\left. \left.  m_{3/2}
\left((N_c-1)\tY_{\abg}-R^{\delta}_{(\alpha}
\tY_{\beta\gamma )\delta}\right) \right] \vphantom{\loopff} \right\}
\ ,
\label{F9}
\end{eqnarray}
where $m_{3/2}=e^{G/2}$ and 
$F^m = e^{G/2} {\hat K}^{m{\overline n}} G_{{\overline n}}$
denote the hidden field auxiliary components
for ${\hat K}^{m {\overline{n}}}$ being the inverse of 
${\hat K}_{{\overline {n}} m}\equiv \partial_m\partial_{\overline{n}}\hat{K}$.
We recall that we are using the unit with $M_P=1$, which applies
also for the cut off $\Lambda$.
The  (generalized) Riemann and Ricci tensors that appear in the above 
are given by
\begin{eqnarray}
R_{m{\overline n}\alpha {\overline {\beta}}} &=& 
\partial_m\partial_{\overline{n}}
{\tilde K_{{\alpha}{\overline{\beta}}}}
-\tilde{K}^{\gamma\overline{\delta}}(\partial_m\tilde{K}_{\alpha
\overline{\delta}})
(\partial_{\overline{n}}\tilde{K}_{\gamma\overline{\beta}})  
\ , 
\nonumber \\
R_{{\alpha}{\overline{\beta}}} &=& 
{\hat K}^{m\overline n}
R_{m {\overline n}{\alpha}{\overline{\beta}}} 
-
{\tilde K^{{\gamma}{\overline{\mu}}}}
{\tilde K^{{\nu}{\overline{\delta}}}}
{Z_{{\alpha}\gamma{\overline{\delta}}}}
{Z_{{\overline{\mu}}{\overline{\beta}}\nu}}
\nonumber \\
&+& 
{\tilde K^{{\gamma}{\overline{\delta}}}}
{\tilde K_{{\alpha}{\overline{\beta}}\gamma{\overline{\delta}}}}
-\tilde{K}^{\gamma\overline{\delta}}\hat{K}^{m\overline{n}}
\partial_{\overline{n}}
Z_{\alpha\gamma}\partial_m Z_{\overline{\beta}\overline{\delta}}\ ,
\nonumber
\\ 
R_{m{\overline{n}}} &=&
\partial_m
\partial_{\overline{n}}\ln [{\rm det}(K_{p\overline{q}})/
{\rm det}({\rm Re} f_{ab})]
+
{\tilde K}^{\gamma\overline{\delta}}
R_{m {\overline n}{\gamma}{\overline{\delta}}} 
\ ,
\nonumber \\
R_m^l &=& (R_{\overline{m}}^{\overline{l}})^*= 
{\hat K}^{l\overline n}
R_{m{\overline{n}}} \ , \quad R_{\alpha}^{\gamma}=\tilde{K}^{\gamma\overline{
\beta}}R_{\alpha\overline{\beta}} \ ,
\label{ricci}
\end{eqnarray}
where ${\tilde K}^{{\overline{\beta}} \alpha}$ is the inverse of 
${\tilde K}_{\alpha {\overline {\beta}}}$, 
${Z_{{\overline{\alpha}}{\overline{\beta}}\gamma}}\equiv
({Z_{{\alpha}\beta{\overline{\gamma}}}})^*$
and 
$Z_{\overline{\alpha}\overline{\beta}}\equiv
(Z_{\alpha\beta})^*$.
The symmetrized subscript in (\ref{F9}) means a cyclic sum,
i.e.  $(\alpha\beta\gamma)=
\alpha\beta\gamma+\beta\gamma\alpha+\gamma\alpha\beta$ and
the K\"ahler covariant derivative $D_m=(D_{\overline m})^*$ is defined  
as 
$D_m H_{\alpha_1...\alpha_p{\overline{\beta}}_1...{\overline{\beta}}_q}= 
\partial_m H_{\alpha_1...\alpha_p{\overline{\beta}}_1...{\overline{\beta}}_q}
-\Gamma_{m\alpha_1}^{\alpha'_1}
H_{\alpha'_1...\alpha_p{\overline{\beta}}_1...{\overline{\beta}}_q}-
...-\Gamma_{m\alpha_p}^{\alpha'_p}H_{\alpha_1...\alpha'_p
{\overline{\beta}}_1...{\overline{\beta}}_q}$
for the K\"ahler connection  
$\Gamma_{m\alpha}^{\gamma}=
{\tilde K^{{ \gamma} {\overline{\rho}} }}
\partial_m {\tilde K_{{\overline{\rho}}{ \alpha}}}$.
(Similarly, $D_{\overline{m}}$ includes 
only  $\Gamma_{\overline{m}\hspace{0.05 cm}\overline{\alpha}}^{\overline{\gamma}}=
(\Gamma_{m\alpha}^{\gamma})^*$ which applies only for 
$\overline{\beta_1},...\overline{\beta_q}$.)
Finally, $V_1$ in the soft scalar masses (\ref{F8}) denotes the vacuum energy
density 
including the quadratically--divergent
one--loop correction:
\begin{eqnarray}
V_1 &&=
\left({\overline {F}}^{{\overline m}}{\hat K}_{{\overline m}n} F^n - 
3m_{3/2}^2\right) \left(1+
\frac{(N_c-5){\Lambda}^2}{16\pi^2}
\right) 
\nonumber\\
&&+\
\frac{{\Lambda}^2}{16\pi^2}
\left[
m_{3/2}^2 (N_c-1) -  
F^{m}
R_{m\overline{n}}
{\overline F}^{\overline n}\hspace{0.05 cm} \right]
\ .
\label{F14}
\end{eqnarray}
The observed vanishing  cosmological constant implies 
that one can set   $V_1=0$ \cite{foot3}.

In (\ref{F8}) and (\ref{F9}), 
the pieces proportional to $\Lambda^2$ correspond
to the one-loop SUGRA modification of the 
matching conditions for the soft parameters at $M_P$. 
These SUGRA corrections appear to be sensitive
to $\Lambda$ whose precise value can be determined
only when the underlying theory of the SUGRA model
is known. 
In the case
that the underlying theory is a string theory,
$\Lambda$ would be either
of order the string scale $M_{\rm st}=\frac{1}{2}g_{GUT}M_P$ 
or of order the compactification
scale $M_c$  which is very close to $M_{\rm st}$
in weakly--coupled heterotic string theory \cite{kaplu}.
In the $M$--theory limit of strong string coupling,
$M_c$ would be lower, but it was argued that
generically $M_c\ger \alpha_{GUT} \sqrt{8\pi}M_P$
\cite{caceres} and
thus $M_c$ is around $M_{\rm st}$ even
in the $M$--theory limit.
Based on these observations, in the following
we will use $\Lambda=\frac{1}{2}g_{GUT}M_P$
when we estimate the size of SUGRA corrections.
As can be easily noted,  SUGRA corrections
involve the summation over the chiral multiplets.
For instance, $R_{\alpha\overline{\beta}}$ 
in (\ref{ricci})
involves  the summation 
of the coefficients 
$R_{m {\overline n}{\alpha}{\overline{\beta}}}$ and
$\tilde{K}_{\alpha\overline{\beta}\gamma\overline{\delta}}$.
Similarly $R_{m\overline{n}}$ involves 
the summation of the coefficients
$R_{m {\overline n}{\alpha}{\overline{\beta}}}$.
Since these coefficients
are  generically of order one in the unit with $M_P=1$,
the SUGRA one-loop corrections in the soft scalar
mass and $A$ are generically
${\cal O}(\frac{N_c\Lambda^2}{16\pi^2 M_P^2}m_{3/2})$
{\it unless} some cancellations take place.
(See also the pieces in (\ref{F8}) and (\ref{F9})
which include explicitly the factor $N_c$. 
It is worth noticing that the gauge kinetic function
piece in $R_{m\overline{n}}$  can give a contribution
of ${\cal O}(\frac{N_v\Lambda^2}{16\pi^2 M_P^2}m_{3/2})$
where $N_v$ is the total number of vector multiplets.)
Since $N_c$ can be large as ${\cal O}(10^2)$
($N_c> 49$ since the minimal supersymmetric standard model 
contains already 49 observable chiral multiplets),
these SUGRA corrections are expected to be quite significant,
for instance 
${\cal O}(10 \, \%)$
corrections
for  $\Lambda=\frac{1}{2} g_{GUT} M_P$ \cite{foot2}.
They would be  particularly
important  if 
$m_{3/2}^2\tilde{K}_{\alpha\overline{\beta}}=F^mR_{
m\overline{n}\alpha\overline{\beta}}\overline{F}^{\overline{n}}$ and thus
the tree--level matching conditions, i.e. setting $\Lambda=0$ in (\ref{F8}), 
give vanishing soft scalar masses.
This  indeed happens in no--scale SUGRA models and also in some special limits
of superstring effective SUGRA models which will be discussed later. 
In this case, the major part of soft scalar masses may arise from
the one-loop SUGRA effects in (\ref{F8}).

One of the interesting features
of the above SUGRA corrections 
is the {\it lack
of universality}. 
If the Riemann tensor can be factorized as
$R_{m {\overline n}{\alpha}{\overline{\beta}}}=c_{m\overline{n}}\tilde{K}_{\alpha
\overline{\beta}}$, as it happens e.g. in the dilaton--dominated
SUSY breaking in string theory or in no--scale SUGRA models,
the tree--level matching conditions would give a universal
soft scalar mass for the {\it normalized} observable fields
with canonical kinetic terms.
But now we see that due to
SUGRA radiative corrections, particularly due to the pieces
depending upon 
${Z_{{\alpha}\beta{\overline{\gamma}}}}$ and 
${\tilde K_{{\alpha}{\overline{\beta}}\gamma{\overline{\delta}}}}$
in (\ref{ricci}),
the soft scalar masses will have a generic matrix structure with 
non--degenerate eigenvalues \cite{nir}. 
So, even when tree--level soft masses are universal,
FCNC effects may appear as a consequence of this modification
of the matching condition.
Note that this is completely independent of the FCNC
effects arising from the running of soft masses induced by
the Yukawa interactions.
Of course, if soft masses are already non-universal at tree level,
SUGRA corrections are an extra source of non-universality to 
be added.

Let us now discuss the  $\mu$ term and the related $B$ coefficient.
The effective superpotential (after the hidden sector fields are
integrated out) will include
the $\mu$ term:
$W_{\rm eff}\ni \frac{1}{2}\mu^{\prime}_{\alpha\beta}C^{\alpha}C^{\beta}$.
(Again the prime in $\mu$ means  that it 
is defined for {\it un-normalized} $C^{\alpha}$
in (4).)
The  {\it supersymmetric} scalar masses  
resulting from this  $\mu$-term
are given by 
$K^{\prime\gamma\overline{\delta}}\mu^{\prime}_{\alpha\gamma}
\mu^{\prime *}_{\overline{\beta}\hspace{0.05cm}\overline{\delta}}$
where $K^{\prime\gamma\overline{\delta}}$ is the inverse of 
$K^{\prime}_{\gamma
\overline{\delta}}$ in (\ref{kco}).
We then find
\bea
\mu^{\prime}_{\albe}&&=
\left(e^{\hK/2}
\frac{\hW^*}{|\hW|}\tmu_{\albe}+
m_{3/2} Z_{\albe} - \overline{F}^{\bn} \partial_{\bn} Z_{\albe} \right)
\left(1+\frac{(N_c-7)\Lambda^2}{32\pi^2}\right)
\nonumber \\
&&-\loopf \left\{ \vphantom{ \sum_a g_a^2 M_a^* \tf_{a\albe}}
\overline{F}^{\bn}
\left[\partial_{\bn} (\hK^{p\bq} D_p \partial_{\bq} Z_{\albe})
-R^{\overline{l}}_{\overline{n}}
\partial_{\bl} Z_{\albe}
+\partial_{\bn}\left(\tK^{\gbd}(Z_{\abgbd}-
\tilde{K}^{\rho\overline{\sigma}}Z_{\alpha\rho\overline{\delta}}
Z_{\beta\gamma\overline{\sigma}} )\right)\right] \right.
\nonumber \\
&&+ \left. \sum_a g_a^2 M_a^* \tf_{a\albe} \right\}
\label{mu}
\eea
where $M_a=\frac{1}{2}g_a^2F^m\partial_m\hat{f}_a$ denotes
the $a$-th {\it normalized} gaugino mass
and we have assumed that the gauge kinetic function is
diagonal in gauge indices, i.e. ${f}_{ab}=\delta_{ab}f_a$,
as is the case in all interesting models.

The above result shows that 
the one-loop SUGRA effects 
proportional to $\Lambda^2$ 
do  not only modify the  already known contributions 
from $\tilde{\mu}_{\alpha\beta}$ and $Z_{\alpha\beta}$
in the tree--level matching condition,
but also provide {\it new} sources of the $\mu$ term
depending upon the coefficients
$Z_{\alpha\beta\overline{\gamma}}$ and 
$Z_{\alpha\beta\gamma\overline{\delta}}$ in the K\"ahler potential
and also the coefficients $\tilde{f}_{ab\alpha\beta}=\delta_{ab}\tilde{f}_{
a\alpha\beta}$ in the gauge
kinetic function (\ref{F4}).
Under the natural assumption that these coefficients
are of order one in the unit with $M_P=1$,
the new contributions from 
$Z_{\alpha\beta\overline{\gamma}}$ and 
$Z_{\alpha\beta\gamma\overline{\delta}}$ are
of ${\cal O}(\frac{N_c\Lambda^2}{16\pi^2 M_P^2} m_{3/2})$,
while the contribution  from  
$\tilde{f}_{a\alpha\beta}$ is of ${\cal O}(\frac{N_v\Lambda^2}{16\pi^2
M_P^2} M_a)$, and thus they are naturally
of order the weak scale.
We stress that our mechanism generating $\mu$ from the gauge kinetic function,
i.e.  the last piece of (\ref{mu}), 
is different from 
the mechanism of \cite{AGNT}
which uses  a hidden sector gauge kinetic function $f_h$
as the origin of $\mu$.
The mechanism of \cite{AGNT} applies
only when SUSY is broken by the   
gaugino condensation generating 
a nonperturbative superpotential  $W_{\rm np}\propto
e^{-c f_h}$,
and thus
in our context, it corresponds to
generating the coefficient $\tilde{\mu}_{\alpha\beta}$ in
the SUGRA superpotential. 
The $B$ coefficients related with the above
$\mu^{\prime}_{\alpha\beta}$ are given by
\bea
B_{\albe}^{\prime}&=&V_1 Z_{\albe}+\left\{
e^{\hK/2}\frac{\hW^*}{|\hW|} \left[ F^n(\mu_{\albe}\partial_n\hK+
D_n\mu_{\albe})-m_{3/2}\mu_{\albe}\right]\right.
+2 m^2_{3/2}Z_{\albe}
\nonumber \\
&-&m_{3/2}\left.\vphantom{\frac{\hW^*}{|\hW|}}\overline{F}^{\bn}\partial_{\bn}Z_{\albe}
+m_{3/2}F^nD_nZ_{\albe}-F^m\overline{F}^{\bn}D_m\partial_{\bn}Z_{\albe}\right\}
\loopff
\nonumber \\
&-&\loopf\left\{
e^{\hK/2}\frac{\hW^*}{|\hW|} \left[ 
F^mR_m^n \left(\mu_{\albe}\partial_n\hK+ D_n\mu_{\albe}\right)
+F^m\left(D_mR^{\gamma}_{(\alpha}\right)\mu_{\beta)\gamma}
\right. \right.
\nonumber \\
&-&m_{3/2}\left.\left(
\mu_{\albe}(N_c-1)-R^{\gamma}_{(\alpha}\mu_{\beta ) \gamma}
\right) \right]
+m_{3/2}^2R^{\gamma}_{(\alpha}Z_{\beta ) \gamma}
+m_{3/2}F^m\left[R_m^nD_nZ_{\albe}
+\left(D_mR^{\gamma}_{(\alpha}\right)Z_{\beta ) \gamma}\right]
\nonumber \\
&-&m_{3/2}\overline{F}^{\bm}\left[R_{\bm}^{\bn}\partial_{\bn}Z_{\albe}
-2\partial_{\bm}\left(\hK^{p\bq}D_p\partial_{\bq}Z_{\albe}\right)
+R_{\alpha}^{\gamma}\partial_{\bm}Z_{\beta\gamma}
+R_{\beta}^{\gamma}\partial_{\bm}Z_{\alpha\gamma}\right]
\nonumber \\
&-& F^m\overline{F}^{\bn}\left[R_{\bn}^{\bq}D_m\partial_{\bq}Z_{\albe}
+R_m^pD_p\partial_{\bn}Z_{\albe}
-2D_m\partial_{\bn}\left(\hK^{p\bq}D_p\partial_{\bq}Z_{\albe}\right) \right.
\nonumber \\
&+&\left.\vphantom{\frac{\hW^*}{|\hW|}}\left.
\left(D_mR_{\alpha}^{\gamma}\right)\partial_{\bn}Z_{\beta\gamma}
+\left(D_mR_{\beta}^{\gamma}\right)\partial_{\bn}Z_{\alpha\gamma}
\right]\right\}
\nonumber \\
&-&\loopf \left\{ \overline{F}^{\bn} \left( F^m D_m D_{\bn}
+2m_{3/2}\partial_{\bn}\right)
\left[\tK^{\gamma\bdelta} (Z_{\abg\bdelta}-
\tK^{\rho\bsigma}Z_{\alpha\rho\bdelta}Z_{\beta\gamma\bsigma})\right]
\right\}
\nonumber \\
&+&\loopf \left\{
\sum_a  g_a^2 \left[2 \left|M_a\right|^2\tilde{f}_{a\albe}
-M^*_a F^m D_m\tilde{f}_{a\albe}
-2 m_{3/2}M^*_a\tilde{f}_{a\albe}\right] \right\} \ .
\eea
Notice that the last two pieces proportional to $\Lambda^2$ are
related with the new sources of the $\mu$ term. Similar comments
to those below (\ref{F14}) with respect to the soft masses and 
$A$ coefficients can be applied to the $B$ coefficients.

The canonically normalized
observable fields can be obtained by the transformation
$C^{\alpha}\rightarrow \omega^{1/2}U_{\alpha\beta}C^{\beta}$ under which
the kinetic coefficients 
go to the identity matrix:
$U_{\alpha\gamma}K^{\prime}_{\gamma\overline{\delta}}
U^*_{\overline{\beta}\hspace{0.05 cm} \overline{\delta}}
=\delta_{\alpha\overline{\beta}}$
and  $\omega^{1/2}$   
($\omega=1+
\frac{\Lambda^2}{32\pi^2}(N_c-N_v+1)$) is due
to the Weyl scaling introduced to make
the graviton kinetic term including the SUGRA one-loop
correction to be the canonical form, i.e.
${\frac{1}{2}}\sqrt{g}\omega {\cal R}\rightarrow \frac{1}{2}\sqrt{g}{\cal R}$.
Then the soft and $\mu$ parameters for the {\it normalized}
fields are given by
\bea
&&
m^2_{\alpha\overline{\beta}}=\omega^{-1}
U_{\alpha\gamma}m^{\prime 2}_{\gamma\overline{\delta}}
U^*_{\overline{\beta}\hspace{0.05 cm}\overline{\delta}} \ ,
\nonumber \\
&& H_{\alpha_1 ... \alpha_p} =
\omega^{-d_H/2}U_{{\alpha_1}{\beta_1}}\cdot\cdot\cdot
U_{{\alpha_p}{\beta_p}}H^{\prime}_{\beta_1 ...\beta_p} \ ,
\label{normal}
\eea
where $H_{\alpha_1 ...\alpha_p}
=(A_{\alpha\beta\gamma}, B_{\alpha\beta}, \mu_{\alpha\beta})$ and
$d_H$ is its mass-dimension.
Note that the above transformation (\ref{normal}) 
toward the normalized fields
provides an additional source of non-universality in soft  masses.

The analysis of \cite{gaillard} shows that the kinetic coefficients
of gauge fields do {\it not} receive any
quadratically--divergent SUGRA correction.
Since gaugino masses are  related with the kinetic coefficients
of  superpartner gauge fields,
this implies that the matching condition
for gaugino masses is affected by the
one--loop SUGRA effects {\it only} through the Weyl scaling
$M_a\rightarrow \omega^{-1/2}M_a$.
This completes our discussion of the one--loop
SUGRA modification of the matching conditions
for $\mu$ and the soft parameters
in generic hidden sector SUGRA models.

Let us now apply our general results
to  the case 
of superstring effective SUGRA models \cite{BIM3}
in which 
the dilaton field $S$ and 
the moduli fields $T_i$ 
play the role of SUSY-breaking hidden sector fields.
At string tree level, the gauge kinetic function is simply given by
${f_a} = k_a S$ 
where $k_a$ is the Kac--Moody level of the gauge factor.
Since the K\"ahler potential depends on the compactification scheme,
we will concentrate here on 
(0,2) symmetric Abelian orbifolds with diagonal moduli and matter metrics.
(However our main conclusions will not depend on the particular compactification
scheme used.)
For this  class of models, some coefficients
of  the
K\"ahler potential in (\ref{F4})
have been computed. In particular,
at string tree level
$\hat K = -\log(S+\overline S) - \sum _i \log(T_i+{\overline T}_i)$ and  
${\tilde K}_{\alpha{\overline{\beta}}}=\delta_{\alpha\beta}
\Pi_i(T_i+{\overline T}_i)^{n_{\alpha }^i}$,
where $n_{\alpha }^i$ are the 
modular weights of the matter fields $C^{\alpha }$. 
Although not computed explicitly yet, one can imagine
the following 
modular--invariant form of 
$\tilde{K}_{\alpha\overline{\gamma}\beta\overline{\delta}}\ :$
${\tilde K}_{\alpha{\overline{\gamma}}\beta{\overline{\delta}}}
=\delta_{\alpha\gamma}\delta_{\beta\delta}  
X_{\alpha\beta}\Pi_i
(T_i+{\overline T}_i)^{n_{\alpha}^i}
\Pi_j
(T_j+{\overline T}_j)^{n_{\beta}^j}$,
where
$X_{\alpha\beta}$
are  constant coefficients of order one.
We then find from
(\ref{F8}) and (\ref{normal}) with $U_{\alpha\beta}=
\delta_{\alpha\beta}  
\Pi_i
(T_i+{\overline T}_i)^{-n_{\alpha}^i/2}
[1-\frac{\Lambda^2}{32\pi^2}(\sum_{\gamma}{X}_{\alpha\gamma}-
\sum_i n_{\alpha}^i-2)]$,
the diagonal soft scalar masses for 
{\it normalized} observable fields:
\begin{eqnarray}
{m}^{2}_{\alpha} &&= 
\left(m_{3/2}^2+\sum_i \frac{n_{\alpha}^i}{(T_i+ {\overline{T}_i})^2}
|F^i|^2 \right) \left(1+\frac{(N_c+N_v-7)\Lambda^2}{32\pi^2}
\right) 
\nonumber\\ 
&&- \frac{\Lambda^2}{16\pi^2}
\left[
\left( \sum_{\gamma} 
{X}_{\alpha\gamma}
-\sum_i n_{\alpha}^i \right)
2m_{3/2}^2 
 +  2
\sum_i \frac{n_{\alpha}^i}{(T_i+ {\overline{T}_i})^2}
\left(2-\sum_{\gamma}n_{\gamma}^i\right) |F^i|^2
\right]
\ ,
\label{supermass}
\end{eqnarray}
where $F^i$ denote the VEVs of the moduli auxiliary fields.

One can now see more explicitly some of the interesting features 
of the one--loop SUGRA corrections which were discussed in the framework
of the general formula (\ref{F8}) for the soft scalar masses.
For instance even when
the tree--level matching condition 
in 
(\ref{supermass})
leads to a universal soft mass, which would be 
the case if all $C^{\alpha}$ have the same modular weight
or if all  $F^i=0$ (this corresponds to the case of dilaton-dominated
SUSY breaking\cite{KL,BIM}),
the SUGRA corrections  
depending upon 
$X_{\alpha\gamma}$
are {\it no longer universal} \cite{lounir}.
Also in the limit 
that tree--level soft masses are vanishing,
which can happen in the
moduli--dominated SUSY breaking if SUSY breaking is equally shared
among $T_i$'s and one considers untwisted particles,
SUGRA one--loop effects in (\ref{supermass}) give
$m_{\alpha}^{2}={\cal O}(\frac{N_c\Lambda^2}{16\pi^2 M_P^2} m_{3/2}^2)
\ap 10^{-1}m_{3/2}^2$ for $\Lambda=\frac{1}{2} g_{GUT}M_P$.
We stress that this SUGRA correction can be much more important
than 
the string one--loop effects which 
would give 
$m^{2}_{\alpha}\ap 10^{-3}m_{3/2}^2$
by modifying the dilaton-dependent  term
in $\hat K$ \cite{BIM}.

\bigskip

{\bf Acknowledgments}:
This work is supported in part by 
the Brainpool Program of KOSEF (CM),
Distinguished 
Scholar  Exchange Program of KRF (KC), and  Basic Science 
Research Institutes Program BSRI-97-2434 (KC).


\begin{thebibliography}{99}
\bibitem{foot1} Precisely speaking, $\{C^{\alpha}\}$ stand for all
chiral multiplets with VEV $\ll M_P$, independently of whether
they have a renormalizable 
coupling with the SUSY standard model sector or not.
\bibitem{foot3} This is valid 
up to ignoring the higher--order loop corrections
which are of ${\cal O}(N_cm_{3/2}^2\Lambda^2/(16\pi^2)^2)$ \cite{CKN}.
In fact, we suspect that in the higher--order calculation 
$V_1$ is replaced by  the vacuum energy density
which includes  the higher loop effects and thus
is zero up to electroweak and QCD effects \cite{pre}.
\bibitem{foot2}
The importance of a large  
number of chiral multiplets was first pointed out in the context of
the one-loop correction to the cosmological constant in \cite{CKN}.
\end{thebibliography}
\end{document}